\documentclass[12pt]{article}
\usepackage{amssymb}
\usepackage{graphicx}
\usepackage{setspace}

\begin{document}

\begin{titlepage}
\begin{center}

\vspace*{20mm}

{\LARGE\bf
\begin{spacing}{1.2}
Towards the stabilization of extra dimensions by brane dynamics
\end{spacing}
}

\vspace*{20mm}

{\large
Noriaki Kitazawa
}
\vspace{10mm}

Department of Physics, Tokyo Metropolitan University,\\
Hachioji, Tokyo 192-0397, Japan\\
e-mail: kitazawa@phys.se.tmu.ac.jp

\vspace*{20mm}

\begin{abstract}

\noindent
All the models of elementary particles and their interactions derived from String Theory
 involve a compact six--dimensional internal space.
Its volume and shape should be fixed or stabilized,
 since otherwise massless scalar fields (moduli) reflecting their deformations appear
 in our four--dimensional space--time, with sizable effects on known particles and fields.
We propose a strategy towards stabilizing the compact space
 without fluxes of three--form fields from closed strings.
Our main motivation and goal is to proceed insofar as possible within conventional string world--sheet theory.
As we shall see,
 D-branes with magnetic flux (``magnetized D-branes'') and the forces between them
 can be used to this end.
We investigate here some necessary ingredients:
 open string one--loop vacuum amplitudes between magnetized D-branes,
 magnetized D-branes fixed at orbifold singularities,
 and potential energies among such D-branes in the compact space that
 result from tree--level closed string exchanges.
\end{abstract}

\end{center}
\end{titlepage}

\section{Introduction}
\label{introduction}

Superstring theory in a flat ten--dimensional space--time can provide a framework
 to describe the Elementary Particles and their interactions
 including gravity beyond the Standard Model.
The Heterotic string theory, with its elegant inclusion of gauge symmetry,
 can naturally accommodate models with grand unification of the gauge interactions
 (see for example \cite{Green-Schwarz-Witten}),
 while the structure of Yukawa couplings in the Standard Model
 can find a natural setting in type IIA/IIB superstring theory with D-branes
 (see for example \cite{Cvetic:2001nr,Aldazabal:2000sa}).
Both scenarios, however, rest on a six--dimensional compact internal space
 whose volume and shape should be fixed or stabilized (moduli stabilization).
Aside from the wide arbitrariness in the choice of vacuum,
 this remains a difficult problem in String Theory.

In this paper we focus on the moduli stabilization problem in type IIB theory,
 which has been extensively studied only in the low--energy effective field theory, the type IIB supergravity,
 introducing three--form fluxes from massless modes of the corresponding closed string
 (``flux compactifications'') \cite{Kachru:2003aw,Kachru:2003sx},
 while it is generally not accessible with string methods.
Therefore,
 it appears important to explore insofar as possible the actual predictions of String Theory
 while keeping within the reach of conventional world--sheet theory.

A scenario without three--form fluxes was already proposed in type I theory
 \cite{Antoniadis:2004pp,Antoniadis:2006eu}.
It rests on the magnetic flux of the D9-brane gauge field in orientifold models
 \cite{Sagnotti:1987tw,Pradisi:1988xd,Horava:1989vt,Horava:1989ga,
       Bianchi:1990yu,Bianchi:1990tb,Bianchi:1991eu,Sagnotti:1992qw},
 and the basic idea of this scenario is simple.
Once the distribution of magnetic fluxes on D9-branes
 is fixed by supersymmetry conditions, or minimum energy conditions,
 volume and shape of the compact space can be fixed
 as a result of quantization conditions of magnetic fluxes.
Since magnetic fluxes on D-branes can be analyzed within conventional world--sheet theory,
 this type of scenario grants calculability.
Although it seems difficult to stabilize all moduli by this simple mechanism,
 it is therefore worth exploring this idea further.

\begin{figure}[t]
\centering
\includegraphics[width=140mm]{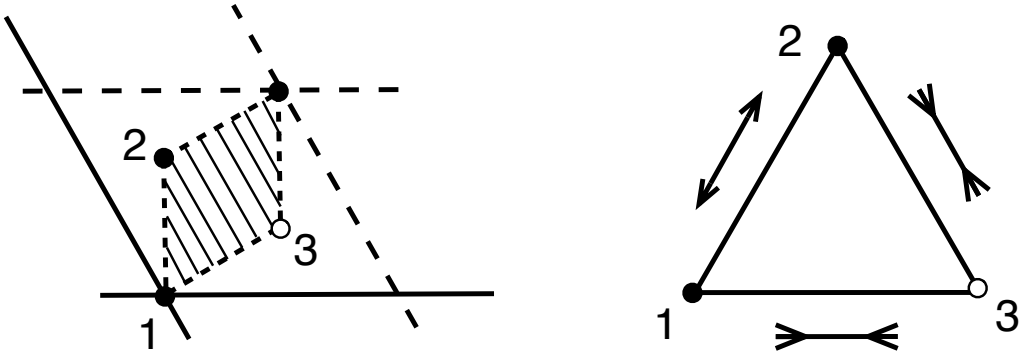}
\caption{
The fundamental region of the $T^2/\bf{Z}_3$ orbifold and the three fixed points (left).
Balance of the forces between three objects located at three different fixed points (right).
As a result, the area of the fundamental region is fixed, or stabilized.
}
\label{fig:balance}
\end{figure}

We investigate the volume stabilization of orbifold compact spaces
 with magnetic fluxes on D-branes in more general situations without supersymmetry.
The D-branes are not necessarily space--time filling, while
supersymmetry may be broken, for instance, by  ``brane supersymmetry breaking''
 \cite{Sugimoto:1999tx,Antoniadis:1999xk,Angelantonj:1999jh,Aldazabal:1999jr,Angelantonj:1999ms}
 while retaining, for the bulk, supersymmetric compactifications.
In addition to the above idea related to magnetic fluxes,
 we examine volume stabilization (K\"ahler moduli stabilization)
 resulting from the balance of attractive and repulsive forces between D-branes in the compact space.
We propose a mechanism that can in principle stabilize the volumes of
 some orbifold spaces with fixed shapes, and thus lacking complex--structure moduli, as for example
 $T^6/{\bf Z}_3$, $T^6/{\bf Z}_7$, $T^6/{\bf Z}_3 \times {\bf Z}_3$
 (see Fig.\ref{fig:balance} for a schematic picture).
In these contexts,
 mutual attractive forces could result from the simultaneous presence of D-branes and anti-D-branes,
 while in principle the non-BPS-branes of \cite{Sen:1998rg,Sen:1998ii,Gaberdiel:1999jd}
 and the fractional non-BPS states of \cite{Dudas:2001wd}
 could provide additional repulsive contributions.
\footnote{
In this paper we do not consider non-geometric dilaton stabilization.
We can naively expect that dilation follows some potential in total system without supersymmetry.
It is well known, for example, that dilaton obtains exponential-type potentials
 in the systems with brane supersymmetry breaking,
 and it could be possible that dilaton follows a racetrack type potential
 with some contributions of branes with negative tension (orientifold fixed planes, for example).
The standard racetrack mechanism with gaugino condensations could also work in our scenario.
We leave this problem to the future work after the achievement of the stabilization of geometric moduli.
}

A concrete model whose compact space is stabilized by this scenario, even if incomplete,
 could allow a related discussion of early Cosmology, and thus of cosmic inflation, within String Theory.
In particular,
 the fields describing brane displacements away from their balanced locations could play the role of inflatons,
 while the vacuum energy of balanced configurations could be the origin of dark energy.
Models of this type
 also possess the attractive feature of linking spontaneous gauge symmetry breaking in String Theory
 to geometrical D-brane displacements, along the lines of \cite{Kitazawa:2012hr}.
One might also conceive of turning the constraint
 that the compact internal space be stabilized, which is unavoidable in a strict sense,
 into a constructive principle to build realistic models of Elementary Particles and their interactions.

In this paper we concentrate on a compact six--dimensional orbifold
 of $T^6/{\bf Z}_3 \times {\bf Z}_3$, which possesses three volume moduli
 (untwisted K\"ahler moduli,
  corresponding to the areas of the three two--tori of $T^6 = T^2 \times T^2 \times T^2$),
 81 blow--up moduli (3 twisted K\"ahler moduli for each of the 27 fixed points)
 and no complex structure moduli.
We investigate the behavior of D$3$-, D$5_1$- and D$7_3$-branes in this compact space,
 where the index identifies the D5-branes whose world--volumes \emph{include} $i$-th torus,
 and the D7-branes whose world--volume \emph{do not include} it.
Since the numbers of Dirichlet--Neumann directions
 of the open string stretched between D$5$- and D$7$-branes are not multiples of 4,
 the system breaks supersymmetry and has a tachyonic ground state.
However, as we shall see the inclusion of appropriate magnetic fluxes on D$7_3$-branes
 can make the system supersymmetric (namely the lowest energy state) while removing the tachyon instability
 (see also the related work of \cite{Angelantonj:2000hi}).
Moreover,
 this configuration fixes the total volume of the first and second tori
 in the $T^6/{\bf Z}_3 \times {\bf Z}_3$ orbifold,
 the inclusion of a D3-brane pose an additional constraint on the magnetic flux on the D$7_3$-brane,
 so that finally the radii of the first and second tori are both fixed.
The D3-brane should be located far enough from the D$5_1$-brane,
 in order to exclude tachyonic ground states for D$3$-D$5_1$ open strings.
The magnetized D$5_1$-D$7_3$ system can be placed
 at a $T^6/{\bf Z}_3 \times {\bf Z}_3$ orbifold fixed point, while
 satisfying twisted Ramond--Ramond tadpole cancelation conditions
 for the consistency at the quantum level.
The simplest system of this type with overall Ramond--Ramond tadpole cancelation
 involves a magnetized D$5_1$-D$7_3$ and its anti--system placed at two different orbifold singularities.
If these are separated in the third torus,
 its radius is driven to shrink by mutual attractive forces,
 so that other objects are needed to stabilize the internal volume.
In this paper we propose a non--trivial treatment of the tadpole problem,
 ubiquitous for D-branes in compact spaces, which can lead to this physically reasonable result.

The paper is organized as follows.
In section \ref{one-loop}
 we provide a brief review of one--loop vacuum amplitudes for open strings ending on magnetized D-branes.
A number of basic facts that are scattered in many articles are collected for later convenience.
In section \ref{stabilization-1}
 the system of magnetized D$5_1$- and D$7_3$-branes is investigated in detail.
The vacuum energy determined by tree--level closed string exchange
 is calculated both in the low--energy effective theory and in the string world--sheet theory.
We shall see in detail that
 the vacuum energy vanishes for a certain configuration of magnetic fluxes,
 when the system possesses supersymmetry.
The resulting configuration of magnetic fluxes
 stabilizes the overall volume of the first and second tori,
 while the inclusion of a D3-brane completes the stabilization of their radii.
In section \ref{singularity}
 we put the system of magnetized D$5_1$- and D$7_3$-branes at a singularity
 of the $T^6/{\bf Z}_3 \times {\bf Z}_3$ orbifold.
The twisted Ramond--Ramond tadpole cancelation
 is non--trivial due to the magnetic fluxes, and supersymmetry is broken.
In section \ref{stabilization-2}
 we discuss the force, or potential energy, between the D$5_1$-D$7_3$ system and its anti--system
 lying at a different singularity separated in third torus,
 and we also propose a non--trivial treatment on the sum of the open--string winding modes in the third torus.
In section \ref{conclusion} we provide a summary of this work and briefly address some future problems.
Many techniques in this paper are familiar to string theorists,
 but we take the freedom to show them in detail for the benefit of others
 who might develop further these ideas in more realistic settings for the purpose of model building.

\section{One--loop vacuum amplitudes on magnetized D-branes}
\label{one-loop}

Let us first investigate in detail the one--loop vacuum amplitude of open strings
 between D$5_1$- and D$7_3$-branes without magnetic flux
 in a $T^6/{\bf Z}_3 \times {\bf Z}_3$ orbifold compactification.
The extension of the arguments to more general configurations should be straightforward.

Among the coordinates of ten--dimensional space--time $X^\mu$ with $\mu = 0, \cdots, 9$
 those of the non--compact four--dimensional space--time bear labels $\mu=0,1,2,3$,
 while those of the compact directions correspond to $\mu = 4,5,6,7,8,9$.
The compact $T^6$ factorizes as $T^2 \times T^2 \times T^2$,
 and the three pairs $\mu = 4,5$, $\mu = 6,7$ and $\mu = 8,9$
 correspond to the first, second and third tori
 with radii $R_1$, $R_2$ and $R_3$, respectively.
We use the SU$(3)$ lattice of Fig.\ref{fig:balance} for all the tori with metric
\begin{equation}
 G_{ab} = \left( \begin{array}{cc} 1 & -1/2 \\ -1/2 & 1 \end{array} \right) \ .
\end{equation}
The twist vectors of the two ${\bf Z}_3$ transformations,
\begin{equation}
 v^{(1)} = (1/3, 0, -1/3) \ ,
 \qquad
 v^{(2)} = (0, 1/3, -1/3) \ ,
\label{twist-vectors}
\end{equation}
specify the angles of discrete rotations in each torus.
Note that the simultaneous action of these two ${\bf Z}_3$
 corresponding to the twist vector $(1/3, 1/3, -2/3)$
 results in a $T^6/{\bf Z}_3$ orbifold that has 9 volume moduli (untwisted K\"ahler moduli),
 while the present $T^6/{\bf Z}_3 \times {\bf Z}_3$ orbifold
 has only 3 volume moduli corresponding to the radii of three tori.
On the other hand, there are 81 twisted K\"ahler moduli in $T^6/{\bf Z}_3 \times {\bf Z}_3$,
 while there are only 27 twisted K\"ahler moduli in the $T^6/{\bf Z}_3$ orbifold.
There are \emph{no} complex structure moduli in both orbifolds.
In this paper
 we do not consider the stabilization of twisted K\"ahler moduli
 against the blow--up of orbifold singularities.

The world--sheet fields of the open string, $X^\mu$ and $\psi^\mu$,
 satisfy Neumann--Neumann boundary condition for $\mu = 0,1,2,3,4,5$,
 Dirichlet--Neumann boundary condition for $\mu = 6,7$,
 and Dirichlet--Dirichlet boundary condition for $\mu = 8,9$.
The open string has Kaluza--Klein modes in the first torus, $\mu = 4,5$,
 and winding modes in the third torus, $\mu = 8,9$.
Although the directions corresponding to $\mu = 6,7$ are compact,
 Dirichlet--Neumann boundary conditions do not allow Kaluza--Klein or winding modes.

The one--loop vacuum amplitude of open strings
 between D$5_1$- and D$7_3$-branes
 separated by a distance $\sqrt{G_{ab} b_a b_b}$ with $a,b=8,9$ in the third torus is
\begin{equation}
 Z_{{\rm D}5 \rightarrow {\rm D}7}^{{\rm no \, flux}}
  = \int_0^\infty {{dt} \over {2t}}
    \sum_{\alpha = 0,1} {1 \over 2} \sum_{\beta=0,1}
    A_{\alpha\beta}^{{\rm D}5 \rightarrow {\rm D}7,{\rm no \, flux}} \ ,
\end{equation}
with
\begin{eqnarray}
 A_{\alpha\beta}^{{\rm D}5 \rightarrow {\rm D}7,{\rm no \, flux}}
 &
 =
 &
  (-1)^\alpha (-1)^{(1-\alpha)\beta}
  {{iV_4} \over {(\sqrt{8 \pi^2 \alpha' t})^4}}
  \sum_{n_4,n_5} q^{{{\alpha'} \over {R_1^2}} G_{ab} n_a n_b}
  \sum_{n_8,n_9}
  q^{{G_{ab} (b_a + 2 \pi R_3 n_a) (b_b + 2 \pi R_3 n_b)} \over {4 \pi^2 \alpha'}}
\nonumber\\
 &
 \times
 &
  {1 \over {(\eta(\tau))^2}}
  \left(
   {{\theta \left[ \begin{array}{c} \alpha/2 \\ \beta/2 \end{array} \right](0,\tau)}
    \over
    {\eta(\tau)}}
  \right)
  \times
  {1 \over {(\eta(\tau))^2}}
  \left(
   {{\theta \left[ \begin{array}{c} \alpha/2 \\ \beta/2 \end{array} \right](0,\tau)}
    \over
    {\eta(\tau)}}
  \right)
\nonumber\\
 &
 \times
 &
  \left(
   {{\theta \left[ \begin{array}{c} \alpha/2-1/2 \\ \beta/2 \end{array} \right](0,\tau)}
    \over
    {\theta \left[ \begin{array}{c} 0 \\ 1/2 \end{array} \right](0,\tau)}}
  \right)
  \times
  {1 \over {(\eta(\tau))^2}}
  \left(
   {{\theta \left[ \begin{array}{c} \alpha/2 \\ \beta/2 \end{array} \right](0,\tau)}
    \over
    {\eta(\tau)}}
  \right) \ ,
\label{D5-D7-no-flux}
\end{eqnarray}
 where $q \equiv \exp(2 \pi i \tau)$ and $\tau \equiv it$.
The sign $(-1)^\alpha$ enforces Fermi statistics in the open--string Ramond sector, while
 the sign $(-1)^{(1-\alpha)\beta}$ determines the Ramond--Ramond charge of the closed string,
 and reflects a non--trivial Gliozzi--Scherk--Olive parity of some Neveu--Schwarz (NS) sectors the open string.
The remaining three factors in the first line of eq.~(\ref{D5-D7-no-flux})
 result from the integration of continuous momenta in the non--compact directions,
 from the summation of Kaluza--Klein momenta in the $4,5$ directions,
 and from the summation of open string winding contributions in the $8,9$ directions.
The four factors in the second and third lines of eq.~(\ref{D5-D7-no-flux})
 represent the contribution of string vibration modes under
 Neumann--Neumann boundary conditions in non--compact directions ($\mu = 0,1,2,3$)
 (the contribution from, say $\mu = 0,1$, are canceled by ghost contributions),
 Neumann--Neumann boundary conditions in the first torus ($\mu = 4,5$),
 Dirichlet--Neumann boundary condition in the second torus ($\mu = 6,7$),
 and Dirichlet--Dirichlet boundary conditions in the third torus ($\mu = 8,9$),
 respectively.
The contribution of vibration modes with Neumann--Neumann boundary condition
 is the same as with Dirichlet--Dirichlet boundary condition,
 and it is different from that with Dirichlet--Neumann boundary condition.
The first systematic understanding of the one--loop open string vacuum amplitudes
 for each combinations of Dirichlet and Neumann boundary conditions
 was given in \cite{Pradisi:1988xd}.

Having defined the building blocks of the open string one--loop amplitude,
 it is simple to understand the one--loop vacuum amplitudes of the open strings
 between D$7_3$- and D$7_3$-branes and D$5_1$- and D$5_1$-branes.
\begin{eqnarray}
 A_{\alpha\beta}^{{\rm D}7-{\rm D}7,{\rm no \, flux}}
 &
 =
 &
  (-1)^\alpha (-1)^{(1-\alpha)\beta}
  {{iV_4} \over {(\sqrt{8 \pi^2 \alpha' t})^4}}
\nonumber\\
 &
 \times
 &
  \sum_{n_4,n_5} q^{{{\alpha'} \over {R_1^2}} G_{ab} n_a n_b}
  \sum_{n_6,n_7} q^{{{\alpha'} \over {R_2^2}} G_{ab} n_a n_b}
  \sum_{n_8,n_9} q^{{{R_3^2} \over {\alpha'}} G_{ab} n_a n_b}
\nonumber\\
 &
 \times
 &
  {1 \over {(\eta(\tau))^8}}
  \left(
   {{\theta \left[ \begin{array}{c} \alpha/2 \\ \beta/2 \end{array} \right](0,\tau)}
    \over
    {\eta(\tau)}}
  \right)^4 \ ,
\label{D7-D7-no-flux}
\end{eqnarray}
\begin{eqnarray}
 A_{\alpha\beta}^{{\rm D}5-{\rm D}5,{\rm no \, flux}}
 &
 =
 &
  (-1)^\alpha (-1)^{(1-\alpha)\beta}
  {{iV_4} \over {(\sqrt{8 \pi^2 \alpha' t})^4}}
\nonumber\\
 &
 \times
 &
  \sum_{n_4,n_5} q^{{{\alpha'} \over {R_1^2}} G_{ab} n_a n_b}
  \sum_{n_6,n_7} q^{{{R_2^2} \over {\alpha'}} G_{ab} n_a n_b}
  \sum_{n_8,n_9} q^{{{R_3^2} \over {\alpha'}} G_{ab} n_a n_b}
\nonumber\\
 &
 \times
 &
  {1 \over {(\eta(\tau))^8}}
  \left(
   {{\theta \left[ \begin{array}{c} \alpha/2 \\ \beta/2 \end{array} \right](0,\tau)}
    \over
    {\eta(\tau)}}
  \right)^4 \ .
\label{D5-D5-no-flux}
\end{eqnarray}

Now we introduce the magnetic flux on D$7_3$-brane in $4,5$ directions (first torus), with $F_{45} > 0$,
 and in $6,7$ directions (second torus), with $F_{67} > 0$.
These magnetic fluxes in compact spaces are quantized as
\begin{eqnarray}
 (2 \pi R_1)^2 \sqrt{{\rm det}G} F_{45} &=& 2 \pi q_1 \ , \qquad q_1 \in {\bf Z} \ ,
\label{quantization_1}
\\
 (2 \pi R_2)^2 \sqrt{{\rm det}G} F_{67} &=& 2 \pi q_2 \ , \qquad q_2 \in {\bf Z} \ ,
\label{quantization_2}
\end{eqnarray}
 where $(2 \pi R_1)^2 \sqrt{{\rm det}G}$ and $(2 \pi R_2)^2 \sqrt{{\rm det}G}$
 are areas of the first and second tori.
For the quantization
 it is convenient to define new world--sheet fields using zweibeins
 so that they form orthonormal bases in the compact space, letting
\begin{equation}
 {\tilde X}^r \equiv e^r{}_a X^a = X^a e_a{}^r \ ,
 \qquad
 {\tilde \psi}^r \equiv e^r{}_a \psi^a = \psi^a e_a{}^r \ ,
\end{equation}
 where
\begin{equation}
 G_{ab} = e_a{}^r {\bf 1}_{rs} e^s{}_b \ ,
\end{equation}
 and concretely
\begin{equation}
 e_a{}^r = \left( \begin{array}{cc} c & s \\ s & c \end{array} \right) \ ,
\end{equation}
 with $c = \cos(\pi/12)$, $s=-\sin(\pi/12)$.
The boundary conditions for the $4,5$ directions on the D$7_3$-brane are
\begin{equation}
 \left\{
 \begin{array}{l}
 (\partial_+ - \partial_-) {\tilde X}^4 + m_1 (\partial_+ + \partial_-) {\tilde X}^5 = 0 \ ,
 \\
 (\partial_+ - \partial_-) {\tilde X}^5 - m_1 (\partial_+ + \partial_-) {\tilde X}^4 = 0 \ ,
 \end{array}
 \right.
 \qquad
 {\rm at} \quad \sigma_1=0,\pi
\label{bc-with-flux-X}
\end{equation}
 with derivatives with respect to world--sheet coordinates $\partial_\pm \equiv (\partial_0 \pm \partial_1)/2$,
 where
\begin{equation}
 m_1 \equiv 2 \pi \alpha' {\tilde F}_{45} \ ,
 \qquad {\tilde F}_{rs} \equiv (e^{-1})_r{}^a F_{ab} (e^{-1})^b{}_s \ ,
\end{equation}
and
\begin{equation}
{\tilde F}_{45} = F_{45}/\sqrt{{\rm det}G} \ .
\end{equation}
Here $\sigma_0$ and $\sigma_1$ are world--sheet coordinates,
 $\partial_{0,1} \equiv \partial/\partial \sigma_{0,1}$,
 and for $m_1=0$ eqs.~(\ref{bc-with-flux-X}) reduce to Neumann--Neumann boundary conditions.
The same happens to world--sheet fermion fields.
\begin{equation}
 \left\{
 \begin{array}{l}
 ({\tilde \psi}_+^4 - e^{ - 2 \pi i \nu} {\tilde \psi}_-^4)
  + m_1 ({\tilde \psi}_+^5 + e^{- 2 \pi i \nu} {\tilde \psi}_-^5) = 0 \ ,
 \\
 ({\tilde \psi}_+^5 - e^{ - 2 \pi i \nu} {\tilde \psi}_-^5)
  - m_1 ({\tilde \psi}_+^4 + e^{- 2 \pi i \nu} {\tilde \psi}_-^4) = 0 \ ,
 \end{array}
 \right.
 \qquad
 {\rm at} \quad \sigma_1=0
\end{equation}
 with $\nu=(1-\alpha)/2$, and
\begin{equation}
 \left\{
 \begin{array}{l}
 ({\tilde \psi}_+^4 - {\tilde \psi}_-^4) + m_1 ({\tilde \psi}_+^5 + {\tilde \psi}_-^5) = 0 \ ,
 \\
 ({\tilde \psi}_+^5 - {\tilde \psi}_-^5) - m_1 ({\tilde \psi}_+^4 + {\tilde \psi}_-^4) = 0 \ ,
 \end{array}
 \right.
 \qquad
 {\rm at} \quad \sigma_1=\pi \ .
\end{equation}
For the D$7_3$-D$7_3$ open string the Virasoro generator $L_0$,
 which determines the spectrum of open string vibrations, is not affected by the magnetic flux,
 but only the quantization condition of momenta in the first torus changes to
\begin{equation}
 p^{4,5} = {1 \over \sqrt{1+m_1^2}} {{n_{4,5}} \over {R_1}} \ ,
\end{equation}
 because
\begin{equation}
 {1 \over \pi} \int_0^\pi d\sigma_1 \ X^{4,5} (\sigma_0=0,\sigma_1) = {1 \over \sqrt{1+m_1^2}} \ x^{4,5} \ ,
\end{equation}
 so that
\begin{equation}
 X^{4,5} \sim X^{4,5} + 2 \pi R_1 \longrightarrow x^{4,5} \sim x^{4,5} + 2 \pi R_1 \sqrt{1+m_1^2} \ .
\end{equation}
This is an interesting result given in \cite{Abouelsaood:1986gd}
 in the case of a constant background $B$-field.
Then we have
\begin{eqnarray}
 A_{\alpha\beta}^{{\rm D}7-{\rm D}7}
 &
 =
 &
  (-1)^\alpha (-1)^{(1-\alpha)\beta}
  {{iV_4} \over {(\sqrt{8 \pi^2 \alpha' t})^4}}
\nonumber\\
 &
 \times
 &
  \sum_{n_4,n_5} q^{{1 \over {1+m_1^2}}{{\alpha'} \over {R_1^2}} G_{ab} n_a n_b}
  \sum_{n_6,n_7} q^{{1 \over {1+m_2^2}}{{\alpha'} \over {R_2^2}} G_{ab} n_a n_b}
  \sum_{n_8,n_9} q^{{{R_3^2} \over {\alpha'}} G_{ab} n_a n_b}
\nonumber\\
 &
 \times
 &
  {1 \over {(\eta(\tau))^8}}
  \left(
   {{\theta \left[ \begin{array}{c} \alpha/2 \\ \beta/2 \end{array} \right](0,\tau)}
    \over
    {\eta(\tau)}}
  \right)^4 \ ,
\label{D7-D7}
\end{eqnarray}
 where $m_2 \equiv 2 \pi \alpha' {\tilde F}_{67}$, and
\begin{eqnarray}
 A_{\alpha\beta}^{{\rm D}5-{\rm D}5}
 &
 =
 &
  (-1)^\alpha (-1)^{(1-\alpha)\beta}
  {{iV_4} \over {(\sqrt{8 \pi^2 \alpha' t})^4}}
\nonumber\\
 &
 \times
 &
  \sum_{n_4,n_5} q^{{{\alpha'} \over {R_1^2}} G_{ab} n_a n_b}
  \sum_{n_6,n_7} q^{{{R_2^2} \over {\alpha'}} G_{ab} n_a n_b}
  \sum_{n_8,n_9} q^{{{R_3^2} \over {\alpha'}} G_{ab} n_a n_b}
\nonumber\\
 &
 \times
 &
  {1 \over {(\eta(\tau))^8}}
  \left(
   {{\theta \left[ \begin{array}{c} \alpha/2 \\ \beta/2 \end{array} \right](0,\tau)}
    \over
    {\eta(\tau)}}
  \right)^4 \ .
\label{D5-D5}
\end{eqnarray}

For the D$5_1$-D$7_3$ open string vacuum amplitude
 the quantization of the string vibration modes is modified by magnetic fluxes,
 because the ends of the open string feel different magnetic fields.
The calculation is straightforward and the result is
\begin{eqnarray}
 A_{\alpha\beta}^{{\rm D}5 \rightarrow {\rm D}7}
 &
 =
 &
  (-1)^\alpha (-1)^{(1-\alpha)\beta}
  {{iV_4} \over {( \sqrt{8 \pi^2 \alpha' t} )^4}}
  \sum_{n_8,n_9}
  q^{{G_{ab} (b_a + 2 \pi R_3 n_a) (b_b + 2 \pi R_3 n_b)} \over {4 \pi^2 \alpha'}}
\nonumber\\
 &
 \times
 &
  {1 \over {(\eta(\tau))^2}}
  \left(
   {{\theta \left[ \begin{array}{c} \alpha/2 \\ \beta/2 \end{array} \right](0,\tau)}
    \over
    {\eta(\tau)}}
  \right)
  \times
  \left(
   {{e^{-i \pi  \beta ({\alpha \over 2} - \alpha_1)}
    \theta \left[ \begin{array}{c} \alpha/2-\alpha_1 \\ \beta/2 \end{array} \right](0,\tau)}
    \over
    {{e^{-i \pi ({1 \over 2} - \alpha_1)}
    \theta \left[ \begin{array}{c} 1/2-\alpha_1 \\ 1/2 \end{array} \right](0,\tau)}}}
  \right)
\nonumber\\
 &
 \times
 &
   \left(
   {{e^{-i \pi  \beta ({\alpha \over 2} - \alpha_2)}
    \theta \left[ \begin{array}{c} \alpha/2-\alpha_2 \\ \beta/2 \end{array} \right](0,\tau)}
    \over
    {{e^{-i \pi ({1 \over 2} -\alpha_2)}
    \theta \left[ \begin{array}{c} 1/2-\alpha_2 \\ 1/2 \end{array} \right](0,\tau)}}}
   \right)
  \times
  {1 \over {(\eta(\tau))^2}}
  \left(
   {{\theta \left[ \begin{array}{c} \alpha/2 \\ \beta/2 \end{array} \right](0,\tau)}
    \over
    {\eta(\tau)}}
  \right) \ ,
\label{D5-D7}
\end{eqnarray}
 where $m_1 = \tan(\pi \alpha_1)$ and $m_2 = \cot(\pi\alpha_2)$ with $0 < \alpha_{1,2} < 1/2$.
Note that there are no Kaluza--Klein modes in the first torus
 because of the magnetic flux on the D$7_3$-brane.
There are many phase factors in the contributions from the $4,5,6,7$ directions
 because of the shift of zero modes by magnetic fluxes on the D$7_3$-brane.
Therefore, the limit of no magnetic fluxes,
 $\alpha_1 \rightarrow 0$ and $\alpha_2 \rightarrow 1/2$,
 does not coincide with the amplitude of no magnetic fluxes.

\section{Volume stabilization by magnetized D-branes}
\label{stabilization-1}

We can now investigate the tree--level closed--string exchange amplitude between D$5_1$- and D$7_3$-branes,
 which is the potential energy between these D-branes, or a contribution to the vacuum energy.

It is instructive to investigate it first in the low--energy effective field theory
 that only includes the massless states of closed string:
 dilaton, graviton, B-field and Ramond--Ramond fields.
For simplicity, we leave aside the effects of compactification.
The couplings of these fields to a D$p$-brane are described
 by the effective action in Einstein frame
\begin{equation}
 S_p = - \tau_p \int d^{p+1} \xi e^{{{p-3} \over 4} \phi}
       \sqrt{-{\rm det} \left( g_{ab} + e^{-{1 \over 2}\phi} (B_{ab} + 2 \pi \alpha' F_{ab}) \right)}
       + i \tau_p \int e^{2 \pi \alpha' F_2 + B_2} \wedge \sum_q C_q \ ,
\end{equation}
 where the integrations are over the D$p$-brane world--volume,
 $\tau_p$ is the D$p$-brane tension,
 $g_{ab}$ and $B_{ab}$ are pull--back tensors of space--time metric and B--field on D$p$-brane,
 $F_{ab}$ is the field strength of the U$(1)$ gauge field on the D$p$-brane.
The differential forms, $F_2$, $B_2$ and $C_q$ ($q=0,2,4,6,8$), in the second term
 refer to the gauge field, to the B-field and to the Ramond--Ramond fields, respectively.
The linear term in each field describes its tadpole coupling to D$p$-brane.
We obtain the propagator of each field in ten--dimensional space--time
 from the type IIB supergravity action in Einstein frame.
The introduction of the magnetic flux, a constant $F_{ab}$, changes tadpole couplings.
For example, the B-field acquires a tadpole coupling with a magnetic flux.

For the D$5_1$-brane
 there are tadpole couplings of graviton, dilaton and Ramond--Ramond fields $C_6$.
For the magnetized D$7_3$-brane in the previous section
 there are tadpole couplings of graviton, dilaton, B-field $B_{45}$ and $B_{67}$
 and Ramond--Ramond fields $C_4$, $C_6$ and $C_8$.
The amplitude resulting from exchanges of
 graviton, dilaton and $C_6$ is in momentum space reads
\begin{equation}
 Z^{\rm SUGRA}
  = i V_6 {{2 \kappa_{10}^2 \tau_5 \tau_7} \over {\vert {\bf k} \vert^2}}
    \left[
     {{(1+m_1^2)m_2^2 + (1 + m_2^2)} \over {2\sqrt{(1+m_1^2)(1+m_2^2)}}}
     - m_2
    \right] \ ,
\end{equation}
 where $m_1$ and $m_2$ are magnetic fluxes defined as in the previous section,
 $\kappa_{10}$ is the gravitational constant in ten--dimensional space--time,
 and ${\bf k}$ is the momentum vector in 8,9 directions of the space.
The first term in the square brackets
 is the contribution of the NS--NS sector,
 while the second term is the contribution of the Ramond--Ramond sector.
The former gives an attractive force while the latter gives a repulsive force
 between D$5_1$- and D$7_3$-branes that can be separated in the 8,9 directions.
Notice that there are solutions of magnetic fluxes that guarantee
the balance of the forces, or a vanishing amplitude:
\begin{equation}
 m_1 = {1 \over {m_2}} \ ,
\label{solution_D5D7}
\end{equation}
 assuming $m_1, m_2 > 0$.
Since the quantity inside square brackets is positive semi--definite,
 the solutions correspond to continuously degenerate vacua.
This simple analysis in the low--energy effective theory is useful
 for estimating the forces between D-branes
 with various constant background B-fields and magnetic fluxes.

After obtaining this result in the low--energy effective theory,
 let us return to the amplitude of eq.~(\ref{D5-D7}).
This amplitude vanishes indeed on account of the identity of eq.~(\ref{jacobi}),
 if $\alpha_1=\alpha_2$, or $m_1=1/m_2$.
\begin{eqnarray}
 \sum_{\alpha,\beta} {1 \over 2}
 A_{\alpha\beta}^{{\rm D}5 \rightarrow {\rm D}7}
 &\propto& \sum_{\alpha,\beta} {1 \over 2} (-1)^\alpha (-1)^{(1-\alpha)\beta}
 \left( \theta \left[ \begin{array}{c} \alpha/2 \\ \beta/2 \end{array} \right](0,\tau) \right)^2
\nonumber\\
 &&
 \times
 e^{-i \pi  \beta ({\alpha \over 2} - \alpha_1)}
    \theta \left[ \begin{array}{c} \alpha/2-\alpha_1 \\ \beta/2 \end{array} \right](0,\tau)
 \,
 e^{-i \pi  \beta ({\alpha \over 2} - \alpha_2)}
    \theta \left[ \begin{array}{c} \alpha/2-\alpha_2 \\ \beta/2 \end{array} \right](0,\tau)
\nonumber\\
 &=&
 q^{\alpha_1^2/2} q^{\alpha_2^2/2} {1 \over 2}
 \sum_{\alpha=0,1} \sum_{\beta = 0,1} (-1)^\alpha (-1)^{(1-\alpha)\beta}
 \left( \theta \left[ \begin{array}{c} \alpha/2 \\ \beta/2 \end{array} \right](0,\tau) \right)^2
\nonumber\\
 &&
 \times
 \theta \left[ \begin{array}{c} \alpha/2 \\ \beta/2 \end{array} \right](-\alpha_1 \tau,\tau)
 \,
 \theta \left[ \begin{array}{c} \alpha/2 \\ \beta/2 \end{array} \right](-\alpha_2 \tau,\tau)
\nonumber\\
 &=&
 q^{\alpha_1^2/2} q^{\alpha_2^2/2}
 \prod_{i=1}^4 \theta \left[ \begin{array}{c} 1/2 \\ 1/2 \end{array} \right](x_i,\tau) = 0 \ ,
\end{eqnarray}
 with
\begin{equation}
 x_1 = {1 \over 2} (- \alpha_1 - \alpha_2) \tau \ ,
 \quad
 x_2 = {1 \over 2} (- \alpha_1 + \alpha_2) \tau \ ,
 \quad
 x_3 = {1 \over 2} (\alpha_1 - \alpha_2) \tau \ ,
 \quad
 x_4 = {1 \over 2} (\alpha_1 + \alpha_2) \tau \ .
\end{equation}
Note that $x_2=0$ and $x_3=0$ for $\alpha_1=\alpha_2$ and
\begin{equation}
 \theta \left[ \begin{array}{c} 1/2 \\ 1/2 \end{array} \right](0,\tau) = 0\ .
\end{equation}
We conclude that a contribution to the vacuum energy by tree--level exchange of closed strings
\begin{equation}
 V_{{\rm D}5 \rightarrow {\rm D}7}
  = - 2 Z_{{\rm D}5 \rightarrow {\rm D}7}
  = - 2 \int_0^\infty {{dt} \over {2t}}
        \sum_{\alpha,\beta} {1 \over 2}
        A_{\alpha\beta}^{{\rm D}5 \rightarrow {\rm D}7}
\end{equation}
 has continuous supersymmetric local minima for magnetic fluxes $\alpha_1=\alpha_2$.
(Here, the factor $2$ is introduced
 to represent the existence of left and right modes of closed string.)
Notice that the quantization conditions of magnetic fluxes,
 eqs.~(\ref{quantization_1}) and (\ref{quantization_2}),
 fix the total volume of the first and second tori, which is proportional to the product $R_1 R_2$.
This is the mechanism of volume stabilization by magnetic fluxes that we had anticipated.
Here, we do not require supersymmetry, but we require that the system should be in a lowest energy state.
Namely, we require that
 the contribution to the energy of this subsystem to that of the total system,
 which is not necessary supersymmetric, should be minimum.

Let us now introduce a D$3$-brane and let us consider the D$3$-D$7_3$ open string vacuum amplitude.
It is exactly the same of eq.~(\ref{D5-D7}),
 except for a difference in the definition of $\alpha_1$: $m_1 = \cot(\pi \alpha_1)$,
 which reflects the change of boundary condition from Neumann--Neumann to Dirichlet--Neumann.
Since the amplitude vanishes for $\alpha_1=\alpha_2$, we have $m_1=m_2$.
Note that $m_1=m_2=0$ is a solution in this case,
 because the combination of D$3$- and D$7_3$-branes is originally supersymmetric
 without magnetic flux.
With the solution of eq.~(\ref{solution_D5D7}),
 the radii of the first and second tori are fixed according to
\begin{equation}
 R_1^2 = {{\alpha' q_1} \over {{\rm det}G}} \ ,
\qquad
 R_2^2 = {{\alpha' q_2} \over {{\rm det}G}} \ .
\end{equation}
The radius of the third torus should be stabilized by some brane dynamics,
 and we shall discuss some possibilities in the remainder of this paper.

Before closing this section let us mention the fate of the open string tachyon excitation
 between D$5_1$- and D$7_3$-branes with magnetic fluxes.
Without magnetic fluxes
 the tachyon state is a Ramond vacuum state (level 0),
 because the vacuum constant in the corresponding Virasoro operator $L_0$ is $-(1-\alpha)/4$.
\begin{equation}
 L_0 \left\vert s_3=-1/2 \right\rangle = 0
 \quad
 \Longrightarrow
 \quad
 \alpha' k^2 - 1/4 = 0
 \quad
 \Longrightarrow
 \quad
 m^2 = -k^2 = -1/4\alpha' \ ,
\end{equation}
 where $s_3$ is the spin in the 6,7 directions.
The fluxes modify the constant in $L_0$ as $-(1-\alpha)(1-\alpha_1-\alpha_2)/2$,
 and the states with lowest level in the NS sector are $1/2-\alpha_1$ and/or $1/2-\alpha_2$.
These are massless states with $\alpha_1=\alpha_2$, and therefore there is no tachyon state.

\section{Magnetized D-branes at orbifold singularities}
\label{singularity}

We need to fix the magnetized D$5_1$-D$7_3$ system at an orbifold singularity
 to stabilize the radius of the third torus, following the idea anticipated
 in section \ref{introduction}.
The system of D$3$-branes at orbifold singularities
 are extensively discussed in \cite{Aldazabal:2000sa}.
The projection operator of the ${\bf Z}_3 \times {\bf Z}_3$ transformation
 should be inserted in the traces of one--loop open string amplitudes.
It is
\begin{eqnarray}
 &&
 {1 \over 3} \left( 1 + \alpha_{(1)} + \alpha_{(1)}^2 \right)
 {1 \over 3} \left( 1 + \alpha_{(2)} + \alpha_{(2)}^2 \right)
\nonumber\\
 &&\quad
  = {1 \over 9}
    \left( 1 + \alpha_{(1)} + \alpha_{(2)}
             + \alpha_{(1)}^2 + \alpha_{(2)}^2 + \alpha_{(1)}\alpha_{(2)}
             + \alpha_{(1)}\alpha_{(2)}^2 + \alpha_{(1)}^2\alpha_{(2)}
             + \alpha_{(1)}^2\alpha_{(1)}^2
    \right) \ ,
\end{eqnarray}
 where operators $\alpha_{(1)}$ and $\alpha_{(2)}$
 generate first and second ${\bf Z}_3$ transformations, respectively.
Since repeating twice a ${\bf Z}_3$ operation is equivalent to the opposite of a single operation,
 the corresponding amplitudes are related by Hermitian conjugation.
Therefore,
 there are four independent twisted sectors corresponding to
\begin{equation}
 \alpha_{(1)} \ , \quad \alpha_{(2)} \ , \quad \alpha_{(1)}\alpha_{(2)} \ , \quad \alpha_{(1)}\alpha_{(2)}^2 \ .
\label{twisted-sectors}
\end{equation}
We specify each twisted sector by two numbers, ($n_{(1)}$, $n_{(2)}$),
 which are powers of $\alpha_{(1)}$ and $\alpha_{(2)}$ resulting in eq.~(\ref{twisted-sectors}),
 from left to right, in $(1,0)$, $(0,1)$, $(1,1)$ and $(1,2)$ sectors.
To achieve Ramond--Ramond tadpole cancelation in each twisted sector,
 we need to introduce multiple D$5_1$- and D$7_3$-branes
 with a non--trivial action of ${\bf Z}_3 \times {\bf Z}_3$ on their Chan--Paton indexes.
The states of the open string between D$5_1$- and D$7_3$-branes
 should carry a Chan--Paton matrix $\lambda_{i_{\rm{D}5} j_{\rm{D}7}}$
 with $i_{\rm{D}5} = 1,2,\cdots,N_{\rm{D}5}$ and $j_{\rm{D}7} = 1,2,\cdots,N_{\rm{D}7}$,
 where $N_{\rm{D}5}$ and $N_{\rm{D}7}$ are numbers of D$5_1$- and D$7_3$-branes
 on an orbifold singularity, respectively.
The simple actions of the first and second ${\bf Z}_3$ on the Chan--Paton matrix are
\begin{equation}
 \lambda \longrightarrow \gamma_{\rm{D}5}^{(1)} \lambda (\gamma_{\rm{D}7}^{(1)})^{-1}
 \qquad
 {\rm and}
 \qquad
 \lambda \longrightarrow \gamma_{\rm{D}5}^{(2)} \lambda (\gamma_{\rm{D}7}^{(2)})^{-1} \ ,
\end{equation}
 respectively.

The $(n_{(1)}, n_{(2)})$ twisted sector
 one--loop vacuum amplitudes of open strings between D$5_1$- and D$7_3$-branes are
\begin{eqnarray}
 \left.
 A_{\alpha\beta}^{{\rm D}5 \rightarrow {\rm D}7}
 \right\vert_{(n_{(1)}, n_{(2)})}
 &
 =
 &
  (-1)^\alpha (-1)^{(1-\alpha)\beta}
  {\rm tr}((\gamma_{\rm{D}5}^{(1)})^{n_{(1)}}(\gamma_{\rm{D}5}^{(2)})^{n_{(2)}})
  {\rm tr}((\gamma_{\rm{D}7}^{(2)})^{-n_{(2)}}(\gamma_{\rm{D}7}^{(1)})^{-n_{(1)}})
\nonumber\\
 &
 \times
 &
  {{iV_4} \over {(\sqrt{8 \pi^2 \alpha' t})^4}}
  \sum_{n_8,n_9}
  q^{{{R_3^2} \over {\alpha'}} G_{ab} n_a n_b \delta_{n_{(1)}+n_{(2)},0 \, {\rm mod} \, 3}}
\nonumber\\
 &
 \times
 &
  {1 \over {(\eta(\tau))^2}}
  \left(
   {{\theta \left[ \begin{array}{c} \alpha/2 \\ \beta/2 \end{array} \right](0,\tau)}
    \over
    {\eta(\tau)}}
  \right)
  \left(
   {{e^{-i 2 \pi ({\alpha \over 2} - \alpha_1) ({\beta \over 2} + v_1)}
    \theta \left[ \begin{array}{c} \alpha/2-\alpha_1 \\ \beta/2 + v_1 \end{array} \right](0,\tau)}
    \over
    {{e^{-i 2 \pi ({1 \over 2} - \alpha_1) ({1 \over 2} + v_1)}
    \theta \left[ \begin{array}{c} 1/2-\alpha_1 \\ 1/2 + v_1 \end{array} \right](0,\tau)}}}
  \right)
\nonumber\\
 &
 \times
 &
   \left(
   {{e^{-i 2 \pi ({\alpha \over 2} - \alpha_2) ({\beta \over 2} + v_2)}
    \theta \left[ \begin{array}{c} \alpha/2-\alpha_2 \\ \beta/2 + v_2 \end{array} \right](0,\tau)}
    \over
    {{e^{-i 2 \pi ({1 \over 2} - \alpha_2) ({1 \over 2} + v_2)}
    \theta \left[ \begin{array}{c} 1/2-\alpha_2 \\ 1/2 + v_2 \end{array} \right](0,\tau)}}}
   \right)
\nonumber\\
 &
 \times
 &
  (-2 \sin \pi v_3)
   {{\theta \left[ \begin{array}{c} \alpha/2 \\ \beta/2 + v_3 \end{array} \right](0,\tau)}
    \over
    {\theta \left[ \begin{array}{c} 1/2 \\ 1/2 + v_3 \end{array} \right](0,\tau)}} \ ,
\label{D5-D7-at-singularity}
\end{eqnarray}
 where
\begin{equation}
 v_i \equiv n_{(1)} v^{(1)}_i + n_{(2)} v^{(2)}_i
\end{equation}
 with twist vectors defined in eq.~(\ref{twist-vectors}).
Here, we set the distance between D$5_1$- and D$7_3$-branes to zero.
\footnote{
In case of $n_{(1)} = n_{(2)} = 0$ this amplitude gives eq.~(\ref{D5-D7}) 
 which is the contribution of the exchange of untwisted closed string (untwisted sector).
The argument of the volume stabilization in previous section applies to this sector.
The effect of twisted sectors
 (the contributions of the exchanges of twisted closed strings)
 to the energy for the volume stabilization will be discussed at the end of this section
 after solving twisted Ramond--Ramond tadpole cancelation conditions.
}
The corresponding amplitudes for open strings
 between  D$7_3$- and D$7_3$-branes and D$5_1$- and D$5_1$-branes read
\begin{eqnarray}
 \left.
 A_{\alpha\beta}^{{\rm D}7-{\rm D}7}
 \right\vert_{(n_{(1)}, n_{(2)})}
 &
 =
 &
  (-1)^\alpha (-1)^{(1-\alpha)\beta}
  {\rm tr}((\gamma_{\rm{D}7}^{(1)})^{n_{(1)}}(\gamma_{\rm{D}7}^{(2)})^{n_{(2)}})
  {\rm tr}((\gamma_{\rm{D}7}^{(2)})^{-n_{(2)}}(\gamma_{\rm{D}7}^{(1)})^{-n_{(1)}})
\nonumber\\
 &
 \times
 &
  {{iV_4} \over {(\sqrt{8 \pi^2 \alpha' t})^4}}
  \sum_{n_4,n_5}
   q^{{1 \over {1+m_1^2}}{{\alpha'} \over {R_1^2}} G_{ab} n_a n_b \delta_{n_{(1)},0 \, {\rm mod} \, 3}}
\nonumber\\
 &
 \times
 &
  \sum_{n_6,n_7}
   q^{{1 \over {1+m_2^2}}{{\alpha'} \over {R_2^2}} G_{ab} n_a n_b \delta_{n_{(2)},0 \, {\rm mod} \, 3}}
  \sum_{n_8,n_9}
  q^{{{R_3^2} \over {\alpha'}} G_{ab} n_a n_b \delta_{n_{(1)}+n_{(2)},0 \, {\rm mod} \, 3}}
\nonumber\\
 &
 \times
 &
  {1 \over {(\eta(\tau))^2}}
   {{\theta \left[ \begin{array}{c} \alpha/2 \\ \beta/2 \end{array} \right](0,\tau)}
    \over
    {\eta(\tau)}}
  \prod_{i=1}^3
  \left[
   (-2 \sin \pi v_i)
   {{\theta \left[ \begin{array}{c} \alpha/2 \\ \beta/2 + v_i \end{array} \right](0,\tau)}
    \over
    {\theta \left[ \begin{array}{c} 1/2 \\ 1/2 + v_i \end{array} \right](0,\tau)}}
  \right] \ ,
\label{D7-D7-at-singularity}
\end{eqnarray}
\begin{eqnarray}
 \left.
 A_{\alpha\beta}^{{\rm D}5-{\rm D}5}
 \right\vert_{(n_{(1)}, n_{(2)})}
 &
 =
 &
  (-1)^\alpha (-1)^{(1-\alpha)\beta}
  {\rm tr}((\gamma_{\rm{D}5}^{(1)})^{n_{(1)}}(\gamma_{\rm{D}5}^{(2)})^{n_{(2)}})
  {\rm tr}((\gamma_{\rm{D}5}^{(2)})^{-n_{(2)}}(\gamma_{\rm{D}5}^{(1)})^{-n_{(1)}})
\nonumber\\
 &
 \times
 &
  {{iV_4} \over {(\sqrt{8 \pi^2 \alpha' t})^4}}
  \sum_{n_4,n_5} q^{{{\alpha'} \over {R_1^2}} G_{ab} n_a n_b \delta_{n_{(1)},0 \, {\rm mod} \, 3}}
\nonumber\\
 &
 \times
 &
  \sum_{n_6,n_7} q^{{{R_2^2} \over {\alpha'}} G_{ab} n_a n_b \delta_{n_{(2)},0 \, {\rm mod} \, 3}}
  \sum_{n_8,n_9}
   q^{{{R_3^2} \over {\alpha'}} G_{ab} n_a n_b \delta_{n_{(1)}+n_{(2)},0 \, {\rm mod} \, 3}}
\nonumber\\
 &
 \times
 &
  {1 \over {(\eta(\tau))^2}}
   {{\theta \left[ \begin{array}{c} \alpha/2 \\ \beta/2 \end{array} \right](0,\tau)}
    \over
    {\eta(\tau)}}
  \prod_{i=1}^3
  \left[
   (-2 \sin \pi v_i)
   {{\theta \left[ \begin{array}{c} \alpha/2 \\ \beta/2 + v_i \end{array} \right](0,\tau)}
    \over
    {\theta \left[ \begin{array}{c} 1/2 \\ 1/2 + v_i \end{array} \right](0,\tau)}}
  \right] \ .
\label{D5-D5-at-singularity}
\end{eqnarray}
Although eqs.~(\ref{D7-D7-at-singularity}) and (\ref{D5-D5-at-singularity})
 vanish by the $\theta$-function identity of Appendix \ref{laplace} with $v_1+v_2+v_3=0$,
 eq.~(\ref{D5-D7-at-singularity}) does not vanish and supersymmetry is broken.

It is well--known that when we translate these amplitudes
 into corresponding tree--level closed string exchanges,
 they acquire an additional factor
 if there are compact directions with Neumann--Neumann boundary condition.
Suppose that
 there are two compact directions with Neumann--Neumann boundary condition parameterized by a complex coordinate $z$.
In the absence of an orbifold twist the
 freely moving open string end returns to the same place after its one--loop motion:
\begin{equation}
 \int dz \langle z \vert z \rangle
 \equiv \lim_{z' \rightarrow z} \int dz \langle z \vert z' \rangle
 = \lim_{z' \rightarrow z} \int dz \delta(z-z') = 1 \ .
\end{equation}
On the other hand, with an orbifold twist the
 open string may come back to the point up to the identification
 of the orbifold transformation generated by $\hat{\alpha}$ with twisted vector $v$:
\begin{equation}
 \int dz \langle z \vert \hat{\alpha} \vert z \rangle
 = \int dz \langle z \vert e^{i 2 \pi v} z \rangle
 = \int dz \delta((1-e^{i 2 \pi v})z)
 = {1 \over {\vert 1-e^{i 2 \pi v} \vert^2}} \int dz \delta(z)
 = {1 \over {(2 \sin \pi v)^2}} \ .
\end{equation}
The physical meaning of this factor
 is to divide by the number of fixed points in the compact directions,
 and it is necessary to consistently obtain the same twisted Ramond--Ramond tadpole cancelation conditions
 in non--magnetized D$3$- and D$7$-branes at an orbifold singularity, for example,
 from D$3$-D$3$ and D$3$-D$7$ amplitudes and from D$7$-D$7$ and D$3$-D$7$ amplitudes.
If $2 \sin \pi v = 0$, this factor should be replaced by unity,
 as should be clear from the preceding arguments.

For the D$5_1$-D$5_1$ amplitude
 the two directions of the first torus bear Neumann--Neumann boundary conditions,
 and the standard factor of $(2 \sin \pi v_1)^{-2}$ is included.
On the other hand,
 for the first torus in the D$5_1$-D$7_3$ amplitude we do not include the factor,
 because the magnetic flux on the D$7_3$-brane modifies open--string boundary conditions.

It is straightforward to obtain twisted tadpole cancelation conditions
 using D$5_1$-D$5_1$ and D$5_1$-D$7_3$ amplitudes.
For the $(1,0)$ twisted sector, since the second torus is untwisted,
 the effect of the winding modes in the second torus can not be forbidden, and
\begin{equation}
 {{\alpha'} \over {R_2^2\sqrt{\rm detG}}} {1 \over \sqrt{3}}
 \, {\rm tr} \left( \gamma_{\rm D5}^{(1)} \right)
 + i e^{-i \pi /3} m_2 \, {\rm tr} \left( \gamma_{\rm D7}^{(1)} \right) = 0 \ .
\end{equation}
Since the phase of the second term can not be produced by powers of
 $\theta \equiv \exp(i 2 \pi /3)$, this condition actually requires
\begin{equation}
 {\rm tr} \left( \gamma_{\rm D5}^{(1)} \right) = 0
 \quad {\rm and} \quad
 {\rm tr} \left( \gamma_{\rm D7}^{(1)} \right) = 0 \ .
\end{equation}
For the $(0,1)$ twisted sector,
 since the first torus is untwisted,
 the effect of Kaluza--Klein modes in the first torus can not be forbidden,
\begin{equation}
 {{R_1^2} \over {\alpha'\sqrt{\rm detG}}} \sqrt{3}
  \, {\rm tr} \left( \gamma_{\rm D5}^{(2)} \right)
 + i e^{-i \pi /3} {1 \over {m_1}} \, {\rm tr} \left( \gamma_{\rm D7}^{(2)} \right) = 0 \ .
\end{equation}
This requires
\begin{equation}
 {\rm tr} \left( \gamma_{\rm D5}^{(2)} \right) = 0
 \quad {\rm and} \quad
 {\rm tr} \left( \gamma_{\rm D7}^{(2)} \right) = 0
\end{equation}
 for the same reasons as above.
For the $(1,1)$ twisted sector
\begin{equation}
 {\rm tr} \left( \gamma_{\rm D5}^{(1)} \gamma_{\rm D5}^{(2)} \right)
 - \theta^2 {\rm tr} \left( \gamma_{\rm D7}^{(1)} \gamma_{\rm D7}^{(2)} \right)
 = 0 \ ,
\end{equation}
 and for the $(1,2)$ twisted sector
\begin{equation}
 {\rm tr} \left( \gamma_{\rm D5}^{(1)} (\gamma_{\rm D5}^{(2)})^2 \right)
 - {\rm tr} \left( \gamma_{\rm D7}^{(1)} (\gamma_{\rm D7}^{(2)})^2 \right)
 = 0 \ .
\end{equation}

Let us discuss two typical solutions of these twisted tadpole cancelation conditions
 and the corresponding massless spectra. The simplest solution is
\begin{equation}
 \gamma_{\rm D5}^{(1)} = \gamma_{\rm D5}^{(2)} = \gamma_{\rm D7}^{(1)} = \gamma_{\rm D7}^{(2)}
 = {\bf 1}_{N \times N}
   \otimes
   \left(
    \begin{array}{ccc}
     1 & & \\
       & \theta & \\
       & & \theta^2
    \end{array}
   \right) \ .
\end{equation}
In the non--compact four--dimensional world--volume of the D$5_1$-brane
 there are the ${\cal N}=1$ gauge multiplet of U$(N)_1 \times$U$(N)_2 \times$U$(N)_3$ gauge symmetry
 and chiral multiplets in the representation $(N_1,N_3^*)$, $(N_2,N_1^*)$ and $(N_3,N_2^*)$,
 which are beautifully represented by a quiver diagram,
 where $N_i$ and $N_i^*$ denote fundamental and anti--fundamental representations
 of the U$(N)_i$ gauge symmetry.
The same structure appears for the D$7_3$-brane.
No massless state arises from the D$5_1$-D$7_3$ open string.

The second solutions is
\begin{equation}
 \gamma_{\rm D5}^{(1)}
 = {\bf 1}_{N \times N}
   \otimes
   \left(
    \begin{array}{ccc}
     1 & & \\
       & \theta & \\
       & & \theta^2
    \end{array}
   \right) \ ,
 \qquad
 \gamma_{\rm D5}^{(2)}
 = {\bf 1}_{N \times N}
   \otimes
   \left(
    \begin{array}{ccc}
     \theta^2 & & \\
              & \theta & \\
              & & 1
    \end{array}
   \right) \ ,
\end{equation}
\begin{equation}
 \gamma_{\rm D7}^{(1)}
 = {\bf 1}_{N \times N}
   \otimes
   \left(
    \begin{array}{ccc}
     1 & & \\
       & \theta & \\
       & & \theta^2
    \end{array}
   \right) \ ,
 \qquad
 \gamma_{\rm D7}^{(2)}
 = {\bf 1}_{N \times N}
   \otimes
   \left(
    \begin{array}{ccc}
     1 & & \\
       & \theta^2 & \\
       & & \theta
    \end{array}
   \right) \ .
\label{gamma-solution}
\end{equation}
In the non--compact four--dimensional world--volume of the D$5_1$-brane
 there is the ${\cal N}=1$ gauge multiplet of U$(N)_1 \times$U$(N)_2 \times$U$(N)_3$ gauge symmetry,
 and there is no chiral multiplet.
The same structure appears for the D$7_3$-brane
 with U$(N)_4 \times$U$(N)_5 \times$U$(N)_6$ gauge symmetry.
The D$5_1$-D$7_3$ open string yields no massless fermion fields,
 but the following massless scalar fields appear:
\begin{center}
\begin{tabular}{cccccc}
U$(N)_1$ & U$(N)_2$ & U$(N)_3$ & U$(N)_4$ & U$(N)_5$ & U$(N)_6$ \\
$N$      & $1$      & $1$      & $N^*$    & $1$      & $1$      \\
$N$      & $1$      & $1$      & $1$      & $N^*$    & $1$      \\
$1$      & $N$      & $1$      & $1$      & $N^*$    & $1$      \\
$1$      & $N$      & $1$      & $1$      & $1$      & $N^*$    \\
$1$      & $1$      & $N$      & $1$      & $1$      & $N^*$    \\
$1$      & $1$      & $N$      & $N^*$    & $1$      & $1$      \\
\end{tabular}
\end{center}
Note that originally the system of D$5_1$- and D$7_3$-branes has no supersymmetry
 without appropriate magnetic fluxes,
 and the amplitude of eq.~(\ref{D5-D7-at-singularity}) does not vanish.

We can now discuss the possibility of D-branes moving away from the singularity
 while keeping ${\bf Z}_3 \times {\bf Z}_3$ invariance.
It may be possible for triples of D$5_1$-branes and/or D$7_3$-branes
 with identification by ${\bf Z}_3 \times {\bf Z}_3$,
 without violating twisted Ramond--Ramond tadpole cancelations
 \cite{Aldazabal:2000sa,Kitazawa:2012hr}.
In the non--compact four--dimensional portions of the world--volumes of such D-branes
 there is an ${\cal N}=2$ gauge multiplet.
For the first solution,
 if the three scalar fields on the $3N$ D$5_1$-branes have the same vacuum expectation values,
 the U$(N)_1 \times$U$(N)_2 \times$U$(N)_3$ gauge symmetry is broken to a single U$(N)$
 with an adjoint chiral multiplet,
 which can form an ${\cal N}=2$ gauge multiplet with a ${\cal N}=1$ U$(N)$ gauge multiplet.
This means that $3N$ D$5_1$-branes may move away from a singularity
 in such a way that three $N$ D$5_1$-branes are identified by ${\bf Z}_3$.
The same is true for the D$7_3$-branes of the first solution.
In the second solution
 there is no such scalar field that suggests D-brane movement away from the singularity.

Note that fixing D-branes at a singularity means stabilization of D-brane moduli,
 since the conditions of twisted tadpole cancelations forbid D-branes
 to move away from the place of the singularity.
The flat directions associated to such motions are not present in the world--volume field theory
 on the D-branes \cite{Aldazabal:2000sa,Blumenhagen:2005tn}.
This is also related with the stabilization of twisted K\"ahler moduli,
 since the vacuum expectation values of twisted K\"ahler moduli fields are related to
 Fayet-Iliopoulos terms on the D-branes.
Once the D-brane moduli are stabilized in the way as above,
 the emergences of Fayet-Iliopoulos terms simply increase the energy of the system
 and the twisted K\"ahler moduli should also be stabilized.
This problem is also related to
 the understanding of spontaneous breaking of gauge symmetry losing its rank \cite{Kitazawa:2012hr}.

Before closing this section
 we discuss the effects of twisted sectors of eq.~(\ref{D5-D7-at-singularity})
 to the stabilization of the radii of the first and second tori.
In case of the solution of eq.(\ref{gamma-solution})
 only the $(1,1)$ twisted sector among four twisted sectors contributes non-trivially.
Since the corresponding amplitude depends on the magnetic fluxes,
 it provides additional potential energy for volume stabilization
 in addition to that from untwisted sector.
Unfortunately,
 the amplitude diverges in the integration over the modulus $t$
 due to the existence of tadpoles in NS-NS sector of twisted closed string.
(There is no tachyon state even though the system is not supersymmetric.)
Therefore,
 without properly applying Fischler-Susskind mechanism \cite{Fischler:1986ci,Fischler:1986tb}
 or tadpole resummations \cite{Dudas:2004nd,Kitazawa:2008hv}
 we can not obtain a definite answer.
The is a difficult generic problem in models without supersymmetry,
 and finding the solution is beyond the scope of this paper.
In this paper we simply assume that the contribution does not disturb the volume stabilization.

\section{D-brane tadpoles in the compact space}
\label{stabilization-2}

We investigate potential energies between D-branes separated in compact spaces.
There is a general problem of tadpole divergence
 due to the contribution of Kaluza--Klein modes and winding modes \cite{Abel:2003ue},
 and this is usually taken to indicate the need for some redefinition of the closed string background.
In the following we interpret the problem differently
 and propose a procedure to obtain potential energies
 with the appropriate periodicities in the compact spaces.
As an example,
 we consider the system of magnetized D$5_1$- and D$7_3$-branes
 at a singularity and its anti--system at a different singularity of
 $T^6/{\bf Z}_3 \times {\bf Z}_3$ orbifold,
 and investigate the potential energy between these branes and anti--branes.

\begin{figure}[t]
\centering
\includegraphics[width=140mm]{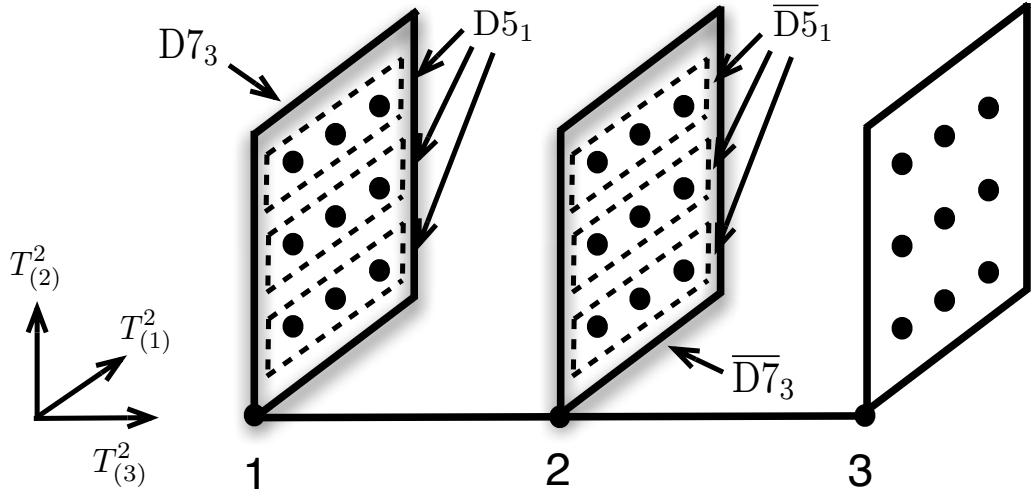}
\caption{
The configuration of magnetized D$5_1$- and D$7_3$-branes
 and its anti-system at $T^6/{\bf Z}_3 \times {\bf Z}_3$ orbifold singularities.
Blobs indicate singularities.
D$7_3$-brane is at the first singularity in the third torus
 and includes nine singularities in its world--volume.
Three D$5_1$-branes are at the first singularity in the third torus
 and at the first, second and third singularities in the second torus, respectively.
Each of the D$5_1$-brane includes three singularities in its world--volume.
At the second singularity in the third torus
 there is the same system with all anti-D-branes.
}
\label{fig:config}
\end{figure}

The D-brane configuration is shown in Fig.\ref{fig:config}.
The system is globally consistent provided all Ramond--Ramond tadpoles cancel,
 and we still have some empty singularities that could host, in principle,
 construct realistic models of elementary particles.
The distance between
 the D-brane system at the first singularity
 and the anti-D-brane system at the second singularity is described by a vector in the third torus.
\begin{equation}
 {\bf b} = b \hat{\bf b}
\end{equation}
 with $b=2 \pi R_3/\sqrt{3}$ and $\hat{\bf b} = (1/\sqrt{3}, 2/\sqrt{3})$.
The potential energy density in four--dimensional space--time
 is a function of $b$, or $R_3$ obtained as follows:
\begin{equation}
 V(R_3) = - 2 {1 \over {iV_4}} \int_0^\infty {{dt} \over {2t}}
        \sum_{\alpha,\beta} {1 \over 2}
        \left[
         A_{\alpha\beta}^{{\rm D}7-\overline{{\rm D}7}}
         + 9 A_{\alpha\beta}^{{\rm D}5-\overline{{\rm D}5}}
         + 3 A_{\alpha\beta}^{{\rm D}7-\overline{{\rm D}5}}
         + 3 A_{\alpha\beta}^{{\rm D}5-\overline{{\rm D}7}}
        \right] \ .
\end{equation}
Here, all one--loop open string amplitudes are those for the untwisted sector,
\begin{eqnarray}
 A_{\alpha\beta}^{{\rm D}7-\overline{{\rm D}7}}
  &=& (-1)^{(1-\alpha)\beta} A_{\alpha\beta}^{{\rm D}7-{\rm D}7} \ ,
\\
 A_{\alpha\beta}^{{\rm D}5-\overline{{\rm D}5}}
  &=& (-1)^{(1-\alpha)\beta} A_{\alpha\beta}^{{\rm D}5-{\rm D}5} \ ,
\\
 A_{\alpha\beta}^{{\rm D}5-\overline{{\rm D}7}}
  &=& A_{\alpha\beta}^{{\rm D}7-\overline{{\rm D}5}}
  = (-1)^{(1-\alpha)\beta} A_{\alpha\beta}^{{\rm D}5 \rightarrow {\rm D}7} \ ,
\end{eqnarray}
 where the amplitudes on the right--hand sides are given in
  eqs.~(\ref{D7-D7}), (\ref{D5-D5}) and (\ref{D5-D7})
  with the introduction of the distance between two D$7_3$-branes and between two D$5_1$-branes.
The potential is described as
\begin{equation}
 V(R_3) =
  - \int_0^\infty {{dt} \over {2t}}
    \sum_{\alpha,\beta}
     V_{\alpha\beta}(t)
    \sum_{n_8,n_9}
     q^{{G_{ab} (b_a + 2 \pi R_3 n_a) (b_b + 2 \pi R_3 n_b)} \over {4 \pi^2 \alpha'}} \ ,
\end{equation}
 where the $V_{\alpha\beta}(t)$ include theta functions
 and the sums of Kaluza--Klein and winding modes in the first and second tori.

One can take the $t \rightarrow 0$ limit (or $s \rightarrow \infty$ limit with $s \equiv \pi/t$)
 to investigate the contributions of massless closed--string modes.
In this limit
\begin{equation}
 \sum_{\alpha,\beta} V_{\alpha\beta}(t) \longrightarrow V_0 \ ,
\end{equation}
 where $V_0$ is a positive constant (provided $R_1$ and $R_2$ are fixed by magnetic fluxes) and
\begin{equation}
 V(R_3) \longrightarrow V(R_3)^{\rm massless} =
  - V_0 \int_0^\infty {{dt} \over {2t}}
    \sum_{n_8,n_9}
     e^{- 2 \pi t \, {{G_{ab} (b_a + 2 \pi R_3 n_a) (b_b + 2 \pi R_3 n_b)} \over {4 \pi^2 \alpha'}}} \ .
\label{potential-massless}
\end{equation}
If we take further the limit using the Poisson summation formula,
\begin{equation}
 V(R_3)^{\rm massless} \longrightarrow
  - V_0 \int_0^\infty {{dt} \over {2t}}
    {{\alpha'} \over {2 R_3^2 \sqrt{{\rm det}G} t}}
 = -V_0 {{\alpha'} \over {2 R_3^2 \sqrt{{\rm det}G}}}
   {1 \over {2 \pi}} \int_0^\infty ds \ ,
\end{equation}
 and this is the ``D$9$-brane'' tadpole divergence that were encountered in \cite{Abel:2003ue}.
This is the general problem
 to investigate the potential energies or forces between D-branes
 that are separated in compact spaces.

Ignoring that 8,9 directions are a compact torus, we have
\begin{equation}
 V(R_3)^{\rm massless} \longrightarrow
  - V_0 \int_0^\infty {{dt} \over {2t}} e^{- {{G_{ab} b_a b_b} \over {2 \pi \alpha'}} t}
 \equiv {{V_0} \over 2} \ln \left( {{G_{ab} b_a b_b} \over {2 \pi \alpha'}} \right) \ ,
\end{equation}
 where we use the argument of Appendix \ref{laplace},
 namely, we define the divergent integral subtracting an infinite constant.
Note that
 a definition of the divergent integral is necessary to obtain
 this physically motivated result.
Therefore,
 it is reasonable to consider a procedure to define the divergent integral
 of eq.~(\ref{potential-massless}).
The summation over the open string winding modes in the third torus
 should translate into the summation of the tree--level propagations of closed string
 winding in the third torus.
Therefore
 it is reasonable to consider eq.~(\ref{potential-massless})
 as an infinite sum of logarithmic potentials that
 preserves torus periodicity.
Following the concrete procedure in Appendix \ref{laplace},
 we obtain the potential energy by massless closed--string exchange:
\begin{equation}
 V(R_3)^{\rm massless}
  \equiv {{V_0} \over 2}
  \left\{
   \ln \left\vert
        {{\theta \left[ \begin{array}{c} 1/2 \\ 1/2 \end{array} \right]
         ({{\tilde b} / {2 \pi R_3}},\tau)}
         \over
         {\eta(\tau)}}
       \right\vert^2
   - {{2 \pi \left( {\rm Im} ({{\tilde b} / {2 \pi R_3}}) \right)^2} \over {{\rm Im} \tau}}
  \right\} \ ,
\end{equation}
 where
\begin{equation}
 \tau \equiv -{1 \over 2} + i {\sqrt{3} \over 2}
 \quad {\rm and} \quad
 \tilde{b} = 2 \pi R_3 \left( {1 \over 3} + \tau \, {2 \over 3} \right) \ .
\end{equation}
Since $\tilde{b}/2 \pi R_3$ is a constant independent of $R_3$,
 the potential energy is also a constant, and the exchanges of massless modes of the closed string
 do not affect the size of the third torus.

The effect of the exchanges of massive modes of the closed string
 can be generally investigated replacing the constant $V_0$
 by an appropriate factor depending on $t$ as
\begin{equation}
 V(R_3)^{\rm massive}_n
  = - a_n \int_0^\infty {{dt} \over {2t}}
     e^{- {{n\pi} \over {2t}}}
     \sum_{n_8,n_9}
     e^{- 2 \pi t \, {{G_{ab} (b_a + 2 \pi R_3 n_a) (b_b + 2 \pi R_3 n_b)} \over {4 \pi^2 \alpha'}}} \ ,
\end{equation}
 with a positive integer $n$.
In the present D-brane system, the coefficient $a_n$ is positive,
the integral is not divergent and
\begin{equation}
 V(R_3)^{\rm massive}_n
  = - a_n \sum_{n_8,n_9}
    K_0\left(
        \sqrt{n \over {\alpha'}}
        \sqrt{{G_{ab} (b_a + 2 \pi R_3 n_a) (b_b + 2 \pi R_3 n_b)}}
        \right) \ ,
\end{equation}
 where $K_\nu(x)$ is a modified Bessel function.
Since the contributions of larger winding modes are exponentially smaller,
 we may approximate this result as
\begin{equation}
 V(R_3)^{\rm massive}_n
  \simeq - a_n K_0\left( \sqrt{n \over {\alpha'}} \sqrt{G_{ab} b_a b_b}\right)
  = -a_n K_0\left( \sqrt{{n \over {\alpha'}}} {{2 \pi R_3} \over \sqrt{3}} \right) \ .
\end{equation}
The factor $\sqrt{n / \alpha'}$ is the mass, because
\begin{equation}
 \int {{d^2k} \over {(2\pi)^2}} {1 \over {k^2+m^2}} e^{-ik \cdot y}
 = {1 \over {2\pi}} K_0(my) \ .
\end{equation}
This potential energy indicates that
 the third torus is forced to shrink by the attractive force between D-brane and anti--D-brane systems,
 although this force decays exponentially with $R_3/\sqrt{\alpha'}$.
For a real stabilization of the third torus,
 it is inevitable to resort to some other objects capable of producing repulsive forces,
 as for instance the non--BPS branes of \cite{Sen:1998rg,Sen:1998ii,Gaberdiel:1999jd}
 and the fractional non--BPS branes of \cite{Dudas:2001wd}.
We leave this analysis to future work.

\section{Conclusions}
\label{conclusion}

We have proposed a strategy
 that can contribute to volume stabilization of internal compact spaces,
 via a combination of magnetic fluxes on D-branes and some D-brane dynamics that all lie
 within the reach of conventional string world--sheet theory,
 and we have explored its features to some extent.
The relevant compact spaces are
 special ones that possess only a small number of K\"ahler moduli (volume moduli)
 but no complex--structure moduli. However, they include some familiar examples,
 since for instance the $T^6/{\bf Z}_3 \times {\bf Z}_3$ orbifold is of this type.
Since we have not fully addressed the stabilization of twisted K\"ahler moduli,
 which correspond to the blow--up of orbifold singularities,
 our scenario is not complete and deserves further investigations.
Nonetheless, we have shown that
 magnetic fluxes on D$7_3$-branes in the presence of non--magnetized D$5_1$- and D$3$-branes can stabilize
 the volume moduli of the first and second tori of the $T^6/{\bf Z}_3 \times {\bf Z}_3$ orbifold,
 as a result of quantization conditions and minimum vacuum energy requirements.

An additional result of this paper concerns
 systems of magnetized D$5_1$-D$7_3$ fractional branes, which
 are fixed at orbifold singularities of the $T^6/{\bf Z}_3 \times {\bf Z}_3$ orbifold, whose
 D$3$ counterparts were discussed in \cite{Aldazabal:2000sa}.
We have also computed the potential energy responsible for the force
 between this D$5_1$-D$7_3$ system and its anti--system lying at different orbifold singularities.
In this computation
 we have proposed a physically motivated prescription to deal with the tadpole divergences
 that generally appear for D-branes in compact spaces,
 which reproduces the expected tendency of the third torus to shrink due to their mutual attractive forces.
It is conceivable that
 repulsive forces introduced by other objects like the non--BPS branes of \cite{Sen:1998rg,Sen:1998ii,Gaberdiel:1999jd}
 and the fractional non--BPS branes of \cite{Dudas:2001wd}
 could stabilize the last volume modulus of $T^6/{\bf Z}_3 \times {\bf Z}_3$ orbifold,
 but we leave a detailed analysis to future work.
The problem of non-geometric dilaton stabilization has not been addressed,
 and we also have to leave this to future work.

\section*{Acknowledgments}

The author would like to thank Augusto Sagnotti for helpful discussions, suggestions
 and a careful reading of the manuscript.
The author also would like to thank Satoshi Iso for encouraging discussions.
This work was supported in part
 by Grant-in-Aid for Scientific Research (\# 26400253) from MEXT Japan,
 by INFN (I.S. Stefi), by the ERC Grant n. 226455 (SUPERFIELDS) and
 by Scuola Normale Superiore.
The author would like to thank Scuola Normale Superiore
 for the kind hospitality extended to him while this work was in progress.

\appendix

\section{Some properties of Theta functions}
\label{theta}

The theta functions used in this paper are defined as
\begin{equation}
 \theta \left[ \begin{array}{c} a \\ b \end{array} \right](z,\tau)
 \equiv e^{2 \pi i a ( z + b)} q^{{1 \over 2}a^2} \prod_{n=1}^\infty ( 1 - q^n )
        \prod_{m=1}^\infty ( 1 + q ^{m + a - {1 \over 2}} e^{2 \pi i (z + b)} )
                           ( 1 + q ^{m - a - {1 \over 2}} e^{- 2 \pi i (z + b)} ) \ ,
\end{equation}
 or
\begin{equation}
 \theta \left[ \begin{array}{c} a \\ b \end{array} \right](z,\tau)
 \equiv \sum_{n=-\infty}^\infty q^{{1 \over 2} (n + a)^2} e^{2 \pi i (n + a) (z + b)} \ ,
\end{equation}
 where $q \equiv {\rm exp}(2 \pi i \tau)$.
This definition is the same as in \cite{Angelantonj:2002ct}. For two integers $n$ and $m$,
\begin{equation}
 \theta \left[ \begin{array}{c} a + n  \\ b + m  \end{array} \right](z,\tau)
 = e^{2 \pi i a n} \theta \left[ \begin{array}{c} a \\ b \end{array} \right](z,\tau) \ ,
\end{equation}
and for $z=0$
\begin{equation}
 \theta \left[ \begin{array}{c} -a \\ -b  \end{array} \right](0,\tau)
 = \theta \left[ \begin{array}{c} a \\ b \end{array} \right](0,\tau) \ .
\end{equation}
It is easy to show that
\begin{equation}
 \theta \left[ \begin{array}{c} a  \\ b  \end{array} \right](z,\tau)
 = \theta \left[ \begin{array}{c} a  \\ b + z  \end{array} \right](0,\tau) \ ,
\end{equation}
 and
\begin{equation}
 \theta \left[ \begin{array}{c} a + \omega  \\ b  \end{array} \right](z,\tau)
 = e^{2 \pi i \omega (z + b)} q^{\omega^2/2}
   \theta \left[ \begin{array}{c} a \\ b \end{array} \right](z + \omega \tau,\tau) \ .
\end{equation}
There famous Riemann identities read
\begin{eqnarray}
 2 &\displaystyle{\prod_{i=1}^4}& \left[ \begin{array}{c} 1/2 \\ 1/2 \end{array} \right](x_i,\tau)
 \nonumber\\
 &=& \prod_{i=1}^4 \left[ \begin{array}{c} 0 \\ 0 \end{array} \right](y_i,\tau)
 - \prod_{i=1}^4 \left[ \begin{array}{c} 0 \\ 1/2 \end{array} \right](y_i,\tau)
 - \prod_{i=1}^4 \left[ \begin{array}{c} 1/2 \\ 0 \end{array} \right](y_i,\tau)
 + \prod_{i=1}^4 \left[ \begin{array}{c} 1/2 \\ 1/2 \end{array} \right](y_i,\tau) \ ,
\label{jacobi}
\end{eqnarray}
 where
\begin{equation}
 \left\{
 \begin{array}{l}
 y_1 = {1 \over 2} ( x_1 + x_2 + x_3 + x_4 ) \ , \\
 y_2 = {1 \over 2} ( x_1 - x_2 - x_3 + x_4 ) \ , \\
 y_3 = {1 \over 2} ( x_1 + x_2 - x_3 - x_4 ) \ , \\
 y_4 = {1 \over 2} ( x_1 - x_2 + x_3 - x_4 ) \ ,
 \end{array}
 \right.
 \qquad {\rm or} \qquad
 \left\{
 \begin{array}{l}
 x_1 = {1 \over 2} ( y_1 + y_2 + y_3 + y_4 ) \ , \\
 x_2 = {1 \over 2} ( y_1 - y_2 + y_3 - y_4 ) \ , \\
 x_3 = {1 \over 2} ( y_1 - y_2 - y_3 + y_4 ) \ , \\
 x_4 = {1 \over 2} ( y_1 + y_2 - y_3 - y_4 ) \ .
 \end{array}
 \right.
\end{equation}
The modular transformation
\begin{equation}
 \theta \left[ \begin{array}{c} a \\ b \end{array} \right](z,\tau)
 = (-i \tau)^{-1/2} e^{2 \pi i a b - i \pi z^2/\tau}
   \theta \left[ \begin{array}{c} -b \\ a \end{array} \right](z/\tau,-1/\tau) \ .
\end{equation}
obtains via a Poisson resummation, and when combined with the Dedekind $\eta$ function
\begin{equation}
 \eta(\tau) \equiv q^{1/24} \prod_{n=1}^\infty (1 - q^n)
 \qquad {\rm with} \qquad
 \eta(\tau) = (-i \tau)^{-1/2} \eta(-1/\tau)
\end{equation}
yields the useful formula
\begin{equation}
 {{\theta \left[ \begin{array}{c} \alpha/2 \\ \beta/2 \end{array} \right](0,\tau)}
  \over
  {\eta(\tau)}}
 =
  {{\theta \left[ \begin{array}{c} \beta/2 \\ \alpha/2 \end{array} \right](0,-1/\tau)}
  \over
  {\eta(-1/\tau)}} \ ,
\end{equation}
 where $\alpha, \beta =0,1$.

\section{Some Green functions in two dimensions}
\label{laplace}

In the calculation of the open string one--loop vacuum energy
 between parallel D$7_i$ and D$7_i$-branes separated by a distance $y = \vert {\bf y} \vert$
 in flat ten--dimensional space--time,
 where ${\bf y}$ is a two--dimensional vector perpendicular to the world--volume of D$7_i$-brane
 with components $y_1$ and $y_2$,
 one encounters the divergent integral
\begin{equation}
 G_2(y) = {1 \over {2 \pi}} \int_0^\infty {{dt} \over {2t}}\ e^{-{{y^2} \over {2 \pi \alpha'}} t} \ ,
\end{equation}
 which is a Green function of the Laplace equation in two--dimensional space, so that
\begin{equation}
 \Delta G_2(y) = - \delta^2(y) \ .
\end{equation}
The integral is defined subtracting an infinite constant using the exponential integral function
\begin{equation}
 E_1(z) \equiv \int_z^\infty dt {1 \over t} e^{-t}
  = -\gamma - \ln(z) - \sum_{k=1}^\infty {{(-z)^k} \over {k k!}} \ ,
\end{equation}
 where $\vert {\rm arg} (z) \vert < \pi$.
Namely, one is resorting to the definition
\begin{equation}
 G_2(y) \equiv {1 \over {2\pi}}
 \lim_{z \rightarrow 0}
 \left[
  \int_z^\infty {{dt} \over {2t}} e^{-{{y^2} \over {2 \pi \alpha'}} t}
  - \int_z^\infty {{dt} \over {2t}} e^{-t}
 \right]
 = - {1 \over {2\pi}} \ln \left( {y \over \sqrt{2\pi\alpha'}} \right) \ .
\end{equation}
This is the standard result for the ``potential energy'' in two--dimensional space.

If one direction perpendicular to D$7_i$-brane world--volume
 ($y_1$ direction) is compact and corresponds to a circle of radius $R$,
 one encounters an infinite summation of logarithmic potentials
 that results from contributions of open string winding states,
\begin{equation}
 G_2^C(y) = -{1 \over {2\pi}} \sum_{n=-\infty}^\infty \ln \sqrt{(y_1 + 2 \pi R n)^2 + (y_2)^2}
          = -{1 \over {2\pi}} \sum_{n=-\infty}^\infty \ln \vert z + 2 \pi R n\vert \ ,
\end{equation}
 where $z \equiv y_1 + i y_2$ with unit $2\pi\alpha'=1$.
This possesses formally circle periodicity, $z \sim z + 2 \pi R$,
 but is divergent one needs to define properly the summation.
Using the product formula
\begin{equation}
 \sin(z) = z \prod_{n=1}^\infty \left( 1 - {{z^2} \over {n^2 \pi^2}} \right)
\end{equation}
 for complex $z$, one can show that
\begin{equation}
 G_2^C(y) = -{1 \over {2\pi}} \ln \vert 2 \sin{{\pi z} \over {2 \pi R}} \vert
            -{1 \over {2\pi}} \ln R -{1 \over {2\pi}} \sum_{n=1}^\infty \ln((2 \pi R n)^2) \ ,
\end{equation}
 and resorting to the analytic continuation of the Riemann zeta function
\begin{equation}
 \zeta(s) \equiv \sum_{n=1}^\infty {1 \over n^s} \ ,
 \qquad \zeta(0) = -{1 \over 2} \ ,
\end{equation}
 finally leads to
\begin{equation}
 G_2^C(y) = -{1 \over {2\pi}} \ln \left\vert 2 \sin{{\pi z} \over {2 \pi R}} \right\vert
            +{1 \over {2\pi}} \left( \ln 2\pi - \sum_{n=1}^\infty \ln((n)^2) \right) \ .
\end{equation}
The first term is the expected Green function of the Laplace equation on a cylinder,
 while the second is an infinite constant that is independent of the radius $R$.
Therefore we define
\begin{equation}
 G_2^C(y) \equiv -{1 \over {2\pi}} \ln \left\vert 2 \sin{{\pi z} \over {2 \pi R}} \right\vert \ ,
\end{equation}
 thus recovering the known Green function of the Laplace equation on a cylinder
 with the corresponding periodicity $z \sim z + 2 \pi R$.

If the two--dimensional space perpendicular to the D$7_i$-brane world--volume
 is compactified on a torus with modulus $\tau$ and radius $R$,
 one is led to a doubly infinite summation of logarithmic potentials,
\begin{equation}
 G_2^T(y) = -{1 \over {2\pi}} \sum_{m=-\infty}^\infty \sum_{n=-\infty}^\infty
             \ln \left\vert z + 2 \pi R \tau m + 2 \pi R n \right\vert \ ,
\end{equation}
Although this formula possesses formally the torus periodicities
 $z \sim z + 2 \pi R$ and $z \sim z + 2 \pi R \tau$,
 the sums are again divergent and one needs to define them appropriately.
First one can apply the definition used in the case of the cylinder to the summation over $n$,
\begin{equation}
 G_2^T(y) \equiv -{1 \over {2\pi}} \sum_{m=-\infty}^\infty
                 \ln \left\vert 2 \sin {{\pi (z + 2 \pi R \tau m)} \over {2 \pi R}} \right\vert
               = -{1 \over {2\pi}} \sum_{m=-\infty}^\infty
                 \ln \left\vert 2 \sin \pi \left( {z \over {2 \pi R}} + \tau m \right) \right\vert \ .
\label{G_T_intermediate}
\end{equation}

The analytic continuation of the Hurwitz zeta function,
 which is commonly used to define the Virasoro generator $L_0$,
\begin{equation}
 \zeta(s,a) = \sum_{n=0}^\infty {1 \over {(a+n)^s}} \ ,
 \qquad \zeta(-1,1) = -{1 \over {12}} \ ,
\end{equation}
leads to the formal relation
\begin{equation}
 0 = {1 \over {12}} {\rm Im} \tau - {1 \over {12}} {\rm Im} \tau
   = {1 \over {12}} {\rm Im} \tau + \sum_{m=1}^\infty m {\rm Im} \tau
   = {1 \over {12}} {\rm Im} \tau
   - {1 \over {2\pi}} \ln \prod_{m=1}^\infty \left\vert e^{2 \pi i \tau m} \right\vert \ ,
\end{equation}
 with which $G_2^T(y)$ can be expressed as a theta function making use of Euler's formula
 for the $\sin$ function in eq.~(\ref{G_T_intermediate}).
The end result is
\begin{equation}
 G_2^T(y) \equiv -{1 \over {2\pi}}
          \ln \left\vert {{\theta \left[ \begin{array}{c} 1/2 \\ 1/2 \end{array} \right](z,\tau)}
                          \over {\eta(\tau)}} \right\vert
          + {{({\rm Im} z)^2} \over {2 {\rm Im} \tau}} \ ,
\label{green-torus}
\end{equation}
 where the second term recovers the torus periodicity that was lost in the manipulations of infinite products.
This is the well--known Green function for the Laplace equation on a torus, such that
\begin{equation}
 \Delta G_2^T(y) \ =\  - \ \delta^2(y) + {1 \over {{\rm Im} \tau}} \ ,
\end{equation}
 where the second term on the right--hand side cancels the charge at ${\bf y}=0$
 in the compact torus space, and is the origin of the second term of eq.~(\ref{green-torus}).

\end{document}